\documentclass[pre,groupedaddress,showkeys,showpacs,twocolumn]{revtex4}
\usepackage{amsmath}
\usepackage{graphics}
\usepackage{graphicx}
\usepackage{amsfonts}
\usepackage{amssymb}
\usepackage{dcolumn}
\usepackage{bm}
\usepackage{natbib}
\begin{document}
\title{Poroelastic indentation of mechanically confined hydrogel layers}
\author{Jessica Delavoipi\`ere}
\author{Emilie Verneuil}
\author{Yvette Tran}
\author{Antoine Chateauminois}
\email[]{antoine.chateauminois@espci.fr}
\affiliation{Soft Matter Science and Engineering Laboratory (SIMM), UMR CNRS 7615,Ecole Sup\'erieure de Physique et Chimie Industrielles (ESPCI), Universit\'e Pierre et Marie Curie, Paris (UPMC), France}
\begin{abstract}
We report on the poroelastic indentation response of hydrogel thin films geometrically confined within contacts with rigid spherical probes of radii in the millimeter range. Poly(PEGMA) (poly(ethylene glycol)) methyl ether methacrylate), poly(DMA) (dimethylacrylamide) and poly(NIPAM) (\textit{N}-isopropylacrylamide) gel films with thickness less than 15 $\mu$m were grafted onto glass substrates using a thiol-ene click chemistry route. Changes in the indentation depth under constant applied load were monitored over time as a function of the film thickness and the radius of curvature of the probe using an interferometric method. In addition, shear properties of the indented films were measured using a lateral contact method. In the case of poly(PEGMA) films, we show that poroelastic indentation behavior is adequately described within the framework of an approximate contact model derived within the limits of confined contact geometries. This model provides simple scaling laws for the characteristic poroelastic time and the equilibrium indentation depth. Conversely, deviations from this model are evidenced for poly(DMA) and poly(NIPAM) films. From lateral contact experiments, these deviations are found to result from strong changes in the shear properties as a result of glass transition (poly(DMA)) or phase separation (poly(NIPAM)) phenomena induced by the drainage of the confined films squeezed between the rigid substrates.
\end{abstract}
\pacs{
     {46.50+d} {Tribology and Mechanical contacts}; 
     {62.20 Qp} {Friction, Tribology and Hardness}
}
\keywords{Friction, rough surfaces, Contact, Rubber, Elastomer, Torsion}
\maketitle
\section*{Introduction}
\indent Polymer gel coatings are suitable candidates for widespread applications in different fields such as medicine and biomedical engineering where their contact mechanical response is a key issue. Polymer gels consist of a network of cross-linked polymer chains swollen with a solvent. Under deformation, the gel behaves as an incompressible material at short times as there is no time for the fluid to flow out of the network. Conversely, the gel acts as a compressible material at extended times as a result of fluid flow under the action of the pressure gradient built up in the fluid phase by the applied stresses. As a consequence, the mechanical response of gels strongly depends on the critical time- and length- scales involved in these transport processes.~\cite{biot1941,hu2012,rice1976,wang2000,hui2005} Within this context, contact experiments between bulk hydrogels and rigid probes were investigated as a tool to determine the elastic properties of gels and the solvent diffusion kinetics within the gels. for that purpose, a theoretical contact mechanics analysis of the indentation of a gel was first derived by Hui~\textit{et al}~\cite{hui2006} where the polymer gel is assimilated to an elastic porous medium and solvent flow is driven by the pressure gradient built up by the applied contact stress. The exact solution derived for the case of a two dimensional Hertz contact between a rigid cylinder and a bulk gel substrate indicates that the initial and final forces required to maintain a given contact size are related to the elastic constants of the network, while the time dependent relaxation of the load is related to the permeability of the network. Accordingly, indentation experiments of bulk hydrogels with rigid probes emerged as a suitable method to probe the elasticity and permeability of gels.~\cite{chan2012a,hu2010,lin2007} When viscoelastic effects come into play, indentation experiments where the contact size is varied were also reported allowing to separate the poroelastic and the viscoelastic response of gels:~\cite{galli2011,hu2010,strange2013} indeed, the poroelastic relaxation time is quadratic in the contact radius while the viscoelastic relaxation time is independent of the contact size.\\
The relevance of poroelastic indentation experiments to the determination of gel properties was further addressed in the case of hydrogel layers lying on rigid substrates. Substrate effects together with the three-dimensional features of stress distribution and solvent flow for low geometric confinements of the film often motivated the use of finite element simulations.~\cite{galli2008,chan2012a,chan2012b,hu2011} As discussed by Galli~\textit{et al}~\cite{galli2008}, the substrate has a twofold effect on the time-dependent contact deformation of the gel: on the one hand, it enhances the hydrostatic pressure and thus the fluid flux beneath the indenter. On the other hand, it constrains the fluid flow along the radial direction. The balance between these competing effects was investigated by Chan and co-workers~\cite{chan2012b} from numerical simulations where the level of confinement of the layers (as quantified from the ratio $\sqrt{R\delta} /e$, where $R$ is the radius of the spherical indenter, $\delta$ is the indentation depth and $e$ is the film thickness) was varied. Under imposed indentation depth conditions, the calculations indicate that the poroelastic relaxation time should decrease with increasing geometric confinements. The validity of this contact mechanics description was supported by a set of indentation experiments carried out by the authors using moderately thin (200-1000~$\mu m$) poly(ethylene glycol)(PEG) hydrogel layers with confinement ratios $\sqrt{R\delta} /e$ ranging from 0.1 to 3: a fit of the data to the numerical model is shown to provide consistent, confinement independent values of the water diffusion coefficient, shear modulus and average pore size of the hydrogel layer.\\
In this study, we report on the poroelastic behaviour of thin (less than 15 $\mu m$) hydrogel layers covalently bonded on a rigid glass substrate under high contact confinement ratios ($\sqrt{R\delta}/e$ ranging from about 10 to 40). We especially address the overlooked issue of the changes in the gel mechanical properties which may arise as a result of the enhanced drainage of the confined films squeezed under high pressure between the rigid glass substrates. Depending on the characteristics of the gel network, the indented film can experience either glass transition or demixing transitions which were not considered in previous studies dealing with moderately confined contacts geometries. For this purpose, different hydrogel networks are considered, namely poly(PEGMA) (poly(ethylene glycol) methyl ether methacrylate), poly(DMA) (dimethylacrylamide) and poly(NIPAM) (N-isopropylacrylamide). While poly(PEGMA) remains in its rubbery state whatever its water content, poly(NIPAM) networks are known to experience either glass transition and/or phase separation phenomena\cite{halperin2015,schild1992} when their water content is lowered. Similarly, poly(DMA) in the dry state has a glass transition temperature (about $110^\circ$~C) well above room temperature.\\
Indentation experiments using a spherical probe are carried out where the changes in the film thickness under a constant applied load are continuously monitored from an interferometric method. We focus on the changes in the poroelastic relaxation time and in the final swelling state of the layers as a function of the film thickness and contact conditions. In addition, time-dependent changes in the mechanical properties of the hydrogel films during the course of poro-elastic indentation are independently determined using a lateral contact method. In a first part, we derive an approximate poroelastic indentation model for coated substrates which is valid within the limits of geometrically confined films. This model allows to derive simple scaling relations describing the dependence of the characteristic poroelastic time and the equilibrium indentation depth on film thickness and contact conditions. Its validity is first investigated for poly(PEGMA) films which are shown to exhibit constant shear properties whatever their water content. Then, we establish the occurrence of glass transition and demixing transitions during the poroelastic indentation of poly(DMA) and poly(NIPAM) films, respectively. These phenomena are evidenced from the time-dependence of their shear properties and from the resulting deviations from the theoretical model which assumes constant shear properties.   
\section*{Materials and methods}
\subsection*{Synthesis of surface-attached hydrogel films}
Poly(NIPAM), poly(DMA) and poly(PEGMA) hydrogel films were synthetized by crosslinking and grafting preformed polymer chains through thiol-ene click chemistry route which is schematically depicted in Supporting Information SI1. This synthesis is preferred to conventional one-shot UV curing of acrylate monomers which are simultaneously polymerized and crosslinked by radical polymerization under controlled atmosphere. Due to high surface to volume ratios, such an UV synthesis of acrylate hydrogel in film form is very sensitive to the presence of oxygen. The problem of ensuring a well controlled environment during polymerization is circumvented by thiol-ene click chemistry.\\
The synthesis of ene-reactive poly(NIPAM) is described in details elsewhere.~\cite{chollet2016,li2015} Briefly, it was carried out in water in two steps: (\textit{i}) free radical polymerization of acrylic acid and \textit{N}-isopropylacrylamide using ammonium persulfate/sodium metabisulfite redox couple as initiator and (\textit{ii}) ene-functionalization of the copolymer by amide formation using allylamine in the presence of EDC/NHS couple. The ene-functionalized poly(NIPAM) was then easily purified by dialysis against water and recovered by freeze-drying.\\
Along the same line, ene-reactive poly(DMA) has been synthesized by free radical polymerization of acrylic acid and dimethylacrylamide followed by ene-functionalization in the same way as for poly(NIPAM). The ene-functionalized poly(DMA) was also purified by dialysis against water and recovered by freeze-drying.\\
For the synthesis of ene-reactive poly(PEGMA), free radical copolymerization of PEGMA ($M_n=300\:g/mol$) and allyl methacrylate (AMA) was performed in toluene using azobisisobutyronitrile (AIBN) as initiator. PEGMA was first purified using an alumina column to remove stabilizer and impurities. The solution of toluene with AIBN and 20 wt$\%$ of PEGMA and AMA was deoxygenated by bubbling nitrogen for 1 hour before thermal activation of AIBN initiator. Radical polymerization was allowed to proceed for 24 hours at 70~$^\circ C$ under nitrogen.\\
For the purpose of thiol-modification of glass substrates, the solid substrates were first cleaned in a freshly prepared ?piranha? (H$_2$SO$_4$/H$_2$O$_2$) solution, rinsed and sonicated in Milli-Q water before drying under nitrogen flow. The freshly cleaned glass substrates were quickly transferred into a sealed reactor filled with nitrogen where a solution of dry toluene with 3 vol$\%$ of 3-mercaptopropyltrimethoxysilane was introduced. After 3 hours immersion in the silane solution under nitrogen, glass substrates were rinsed and sonicated in toluene before drying.\\ 
Surface-attached hydrogel films were synthesized by simultaneously crosslinking and grafting ene-reactive preformed polymer by thiol-ene reaction. The ene-reactive polymer solution was spin-coated on thiol-modified glass substrates with dithioerythritol crosslinkers for poly(NIPAM) and poly(DMA) films and with 2,2'-(ethylenedioxy)diethanethiol crosslinkers for poly(PEGMA) films (with 30-fold molar excess of dithiol to ene-reactive copolymer units). The conditions of spin-coating were fixed with the final angular velocity of 3000 rpm and the spinning time of 30 seconds. Prior to spin-coating, ene-functionalized poly(NIPAM) and dithioerythritol crosslinkers are dissolved in a mixture of butanol and methanol (V/V = 1/1) while ene-functionalized poly(DMA) and dithioerythritol are dissolved in dimethylformamide (DMF). The solution of ene-reactive poly(PEGMA) in toluene was spin-coated as it was (without any purification). After spin-coating, polymer films were annealed at 120~$^\circ C$ for at least 16 hours under vacuum to activate thiol-ene reaction. The glass substrates were then rinsed and sonicated in organic solvent (alcohol for poly(NIPAM), DMF for poly(DMA) and toluene for poly(PEGMA)) and water to remove all free polymer chains (and for poly(PEGMA) residual monomers, initiators and products from the polymer synthesis).\\
The schematics of the syntheses and the characteristics of polymers are provided in the Supporting Information SI2 and SI3. As reported elsewhere,~\cite{chollet2016,li2015} these procedures allow to obtain homogeneous gel films with well controlled mesh sizes and strong adhesion with covalent bonds to the glass substrates. Here, a proof of the efficiency of the chemical bonding of the films to the substrate is provided by their stability during the repeated rinsing steps of the films in various organic solvents and water which are involved in the synthesis. During these operations, the swelling of the films by a factor greater than 2.5 induces strong internal (compressive) stresses which would invariably result in debonding mechanisms at the film/substrate interface unless the film is strongly, covalently bonded to the glass substrate. Accordingly, debonding processes and film removal are systematically observed in solvents when the glass substrate is not thiol-functionalized, i.e. when adhesion at the film/substrate interface only involves physical interactions.\\
The thickness $e_0$ of the swollen hydrogel layers was deduced from profilometry measurements of the thickness of the dried films and from their swelling ratios as they were measured by ellipsometry in water using thin (less than 1~$\mu$m) films (see supporting information SI2 and SI3). These measurements were used as a control of the crosslinks densities of the film networks which are ruled by the ratio of ene-functionnalization. 
\subsection*{Indentation experiments}
\begin{figure}[ht]
	\begin{center}
		\includegraphics[width=0.5\columnwidth]{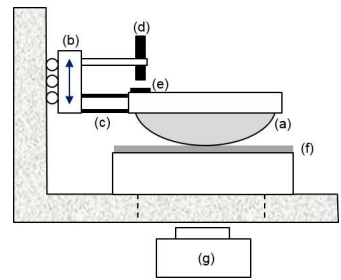}
		\caption{Schematic of the custom indentation set-up. A spherical indenter (a) is fixed to a vertical translation stage (b) by means of a cantilever with two flexible arms (c). During the application of normal contact loading, a measurement of the deflection of the cantilever by means of an optical fiber (d) and a  reflecting surface (e) allows to determine the applied normal load on the film (f). Images of the contact region are recorded with a microscope and a CCD camera (g) operated with monochromatic ($\lambda=546$~nm) light.} 
		\label{fig:setup}
	\end{center}
\end{figure}

\indent Normal contact experiments were carried out where a rigid spherical probe (BK7 glass lens with a radius of curvature of 5.2 or 20.7~mm) is indenting the hydrogel film supported by a glass slide under an imposed applied normal force. A schematic of the set-up is shown in Fig.~\ref{fig:setup}. Normal load control is achieved by means of a double cantilever set to a manual translation stage. The deflection of the blades of the cantilever is continuously monitored using a high precision optical sensor (Philtec, Model D25) and a mirror located on the cantilever tip. It is converted to normal load using the calibrated value of the cantilever stiffness. At the beginning of the indentation experiments, the desired normal load (between 10~mN and 1~N) is applied manually within about $1~s$ using the translation stage. After this loading step, it was verified that the variation in the cantilever deflection resulting from the time-dependent changes in the indentation depth of the poroelastic layer does not induce any significant drop (less than 5\%) of the  normal load during the course of the experiments.\\
The indentation depth is continuously monitored from optical interferences measurements in reflection mode. For that purpose, contact visualization under monochromatic illumination ($\lambda=546\:nm$) is achieved by means of an inverted microscope equipped with a black and white CCD camera (2048x2048 pixels$^2$, 8 bits) operated at frame rates ranging from 1 to 40~Hz. The time-dependent light intensity of the interferences fringes formed between the flat glass substrate and the glass lens allows to determine the changes in the gap between these two surfaces. The indentation depth $\delta$ was taken as the change in this gap from the onset of contact loading (as detected from the sharp increase in load signal). The refractive index of the gel layer $n$ was assumed to be linearly related to the water ($n_w$) and polymer ($n_p$) refractive indexes through water volume fraction $\phi$ : $n=\phi n_w +(1-\phi) n_p$.
\subsection*{Shear measurements}
\indent In order to determine the changes in the mechanical properties of the films during poroelastic indentation, cyclic shear experiments were carried out using a lateral contact method where a small amplitude cyclic shear loading  is superimposed to the applied static normal load. Provided that the displacement is kept low enough (i.e. in the sub-micrometer range), the film can be sheared without significant micro-slip at the contact interface. In such a situation, the contact lateral response thus provides a measurement of the rheology of the polymer film under a constant applied contact load. This method has already been fully described and validated elsewhere using a home-made set-up.~\cite{gacoin2006,janiaud2011} As for the normal indentation experiments, it is based  on a sphere-on-flat contact configuration. During experiments, repeated small amplitude sinusoidal lateral motions are applied to the coated glass substrate (the contacting lens is fixed) by means of a piezoelectric actuator in closed-loop control with a high resolution (better than 10 nm) optical displacement transducer located very close to the contact region in order to get rid of set-up compliance. Lateral force is measured using a high resolution (better than 5 mN) piezoelectric transducer while the constant applied normal force is controlled using a strain gauge load transducer.\\
During each shear cycle, the lateral force and displacement are continuously recorded in order to get the complex lateral contact stiffness defined as $K^{*}=Q^{*}/u$, where $Q^{*}$ is the complex lateral force and $u$ the displacement amplitude at the considered frequency. For the soft hydrogel films under consideration (modulus in the kPa-MPa range), the deformation of the glass substrates can be neglected and the complex viscoelastic shear modulus of the films $G^*$ is thus simply given by $G^*=K^*e/(\pi a^2)$ where $e$ is the film thickness and $a$ is the contact radius. Experiments are carried out using a spherical glass probe with a radius of curvature of $25$~mm. The mismatch between the refractive indices of the water and the swollen hydrogel film was not high enough to allow for a precise determination of the contact radius in this set-up. In the case of experiments carried out in immersion, we therefore only report the measured contact stiffness and the damping factor $\tan \delta$ which is independent on the contact geometry. On the other hand, experiments carried out in air with the dry rubbery poly(PEGMA) films allowed to measure the contact radius $a$ and thus to determine $G'$ and $G"$.\\
\section*{Confined poroelastic contact model}
\begin{figure}[ht]
	\begin{center}
		\includegraphics[width=0.5\columnwidth]{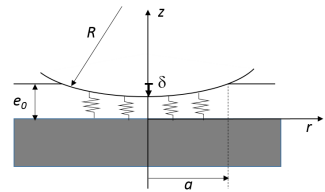}
		\caption{Schematic of the oedometric sphere indentation of an elastic layer lying on a rigid substrate.} 
		\label{fig:contact}
	\end{center}
\end{figure}

\indent In this section, an approximate contact model is derived to describe the poroelastic indentation response of hydrogel films mechanically confined within contact with spherical probes. The contact configuration is schematically depicted in Fig.\ref{fig:contact}. The gel film is assumed to be perfectly bonded to the substrate and the contact between the film and the sphere is considered as frictionless. Both the flat underlying glass substrate and the sphere are taken as perfectly rigid bodies. The model is derived within the limits of confined contact geometries, i.e. conditions where the contact radius $a$ is much greater than the film thickness $e_0$ (or equivalently $\sqrt{R\delta} /e_0>>1$). As a result of such a mechanical confinement, shear deformation in the gel layer is assumed to be negligible, i.e. we only consider vertical displacement components within the film. Along the same line, we assume that film deformation does not expand outside the contact zone.\\
The time-dependent poroelastic response of the hydrogel material is described along the lines of Biot's theory.~\cite{biot1941,biot1955} Accordingly, the hydrogel material is treated as a porous continuum with the pore pressure as a state variable. The constitutive poroelastic response giving the dependence of strain and fluid content on stress and pore pressure is then coupled with Darcy's law for pore fluid transport.\\
At time $t=0$, a normal indentation force $F$ is instantaneously applied to the swollen gel film of initial thickness $e_0$. Considering $R>>\delta$ and $R>>e_0$, the spherical indenter is approximated by a parabola. The vertical displacement $u_z$ within the film can thus be expressed as
\begin{align}
	u_z(z,r,t)&=z\epsilon(r,t) \:\:\: r \leq a\\
	u_z(z,r,t)&=0 \:\:\: r>a \:,
\end{align}
where the vertical deformation $\epsilon(r,t)$ can be written as follows
\begin{equation}
	\epsilon(r,t)=\frac{e_0-e}{e_0}=\frac{\delta(t)-r^2/2R}{e_0} \:,
	\label{eq:epsilon}
\end{equation}
with $\delta$ the indentation depth and $R$ the radius of curvature of the spherical indenter. According to the mixture theory developed by Biot~\cite{biot1955}, the stress is composed of two parts: one generated by the hydrostatic pressure of water in the pores and the other induced by the average stress within the polymer network. The normal stress $\sigma$ within the film thus writes
\begin{equation}
	\sigma(r,t)=\widetilde{E}\epsilon(r,t)+p \:,
	\label{eq:sigma}
\end{equation}
where $p$ is the hydrostatic pressure within the pores and $\widetilde{E}$ is the uni-axial (or oedometric) compression modulus of the fully drained polymer network given by
\begin{equation}
	\widetilde{E}=\frac{2G\left(1-\nu\right)} {1-2\nu}\:,
	\label{eq:E_oedometric}
\end{equation}
where $\nu$ is the Poisson's ratio of the drained polymer network and $G$ is its shear modulus. In eq.~(\ref{eq:sigma}), we have used the convention where a positive pore pressure corresponds to hydrostatic tension. It can be noted that the first term in the rhs of eq.~(\ref{eq:sigma}) corresponds to the formulation of a Winkler elastic foundation model \cite{Johnson1985} where the oedometric nature of the modulus is accounted for.\\
Volume conservation of the liquid in the gel film can be written as
\begin{equation}
	\frac{\partial \epsilon}{\partial t}+\overrightarrow{\nabla}.\overrightarrow{J} =0 \: ,
	\label{eq:volume_conservation}
\end{equation}
where $\overrightarrow{J}$ is the liquid flux.
The radial component of the flux $J_r$ of the liquid within the porous polymer network is related to the pressure gradient by Darcy's law
\begin{equation}
	J_r=-\kappa \frac{dp}{dr} \: ,
	\label{eq:darcy}
\end{equation}
with $\kappa=D_p/\eta$ where $D_p$ is the permeability of the polymer network and $\eta$ is the viscosity of the solvent. $\kappa$ is assumed to be uniform. Due to geometric confinement ($e_0<<a$), lubrication approximation holds and $J_r$ is the only non-vanishing component of the flux, insertion of eq.~(\ref{eq:darcy}) in eq.~(\ref{eq:volume_conservation}) gives
\begin{equation}
	\frac{\partial \epsilon}{\partial t}=-\kappa\frac{1}{r}\frac{\partial}{\partial r }\left(r \frac{\partial p}{\partial r}\right ) \:\: r \leq a.
	\label{eq:epsilonpoint}
\end{equation}
From eq.~(\ref{eq:epsilon}) and (\ref{eq:epsilonpoint}), the relationship between indentation rate $\dot{\delta}$ and pressure gradient is expressed as
\begin{equation}
	\dot{\delta}\left( t \right)=-e_0\kappa \frac{1}{r} \frac{\partial}{\partial r}  \left( r \frac{\partial p}{\partial r} \right)\: ,
	\label{eq:displacement_rate}
\end{equation}
where $(\dot{})$ stands for time derivative. The above differential equation can be solved taking as a boundary condition a zero pore pressure at $r=a$ at all times. The pore pressure profile is then described by the following parabolic distribution
\begin{equation}
	p(r,t)=-\frac{\dot{\delta}}{e_0}\frac{1}{4\kappa}\left(r^2-a^2\right).
	\label{eq:pressure}
\end{equation}
The applied force $F$ can be derived from the integration of the normal stress $\sigma(r,t)$ given by eq.~(\ref{eq:sigma})
\begin{equation}
	F=\int_{0}^{a(t)}dr\sigma2\pi r =2\pi\int_{0}^{a(t)}drr\left( \tilde{E}\epsilon +p \right ).
	\label{eq:force}
\end{equation}
Substituting eq.~(\ref{eq:epsilon}) and (\ref{eq:pressure}) in eq.~(\ref{eq:force}) and using $a(t)=\sqrt{R\delta}$ yields
\begin{equation}
	\delta^2 \left( \tilde{E}+\frac{R\dot{\delta}}{2\kappa}\right )=\frac{Fe_0}{\pi R}.
	\label{eq:equadiff}
\end{equation}
When $t\rightarrow \infty $, $\dot{\delta}=0$ and the final value of the indentation depth is
\begin{equation}
	\delta_{\infty}=\left[ \frac{Fe_0}{\pi R \tilde{E}}\right]^{1/2}\:.
	\label{eq:delta_inf}
\end{equation}
It can be noted that $\delta_{\infty}$ in eq.~(\ref{eq:delta_inf}) is unbounded as $F$ increases. Physically, the maximum achievable indentation depth will obviously be limited by the thickness of the fully drained, dry, polymer film. Eq.~(\ref{eq:equadiff}) can be rewritten in a non dimensional form as
\begin{equation}
	\Delta^2 \left(1+\tau \dot{\Delta}\right)=1\:,
	\label{eq:equadiff_nondim}
\end{equation}
where $\Delta=\delta/\delta_{\infty}$ and $\tau$ is the characteristic poroelastic time defined by
\begin{equation}
	\tau=\frac{\eta}{2D_p} \left[\frac{Re_0F}{\pi\tilde{E}^3}\right]^{1/2}\: .
	\label{eq:tau}
\end{equation}
The differential equation~(\ref{eq:equadiff_nondim}) can be integrated with $T=\frac{t}{\tau}$ yielding
\begin{equation}
	T=-\Delta+\frac{1}{2}Log\left( \frac{1+\Delta}{1-\Delta}\right) \: .
	\label{eq:sol}
\end{equation}

At short times, i.e. when $\Delta<<1$, the time dependence of the normalized indentation depth obeys asymptotically a power law:
\begin{equation}
	\Delta \sim T^{1/3}\: ,
\end{equation}
while at long times it can be approximated by
\begin{equation}
	\Delta \sim 1-e^{-2T}\: ,
\end{equation}
\section*{Poroelastic response of poly(PEGMA) films}
\begin{figure}
	\begin{center}
		\includegraphics[width=0.9\columnwidth]{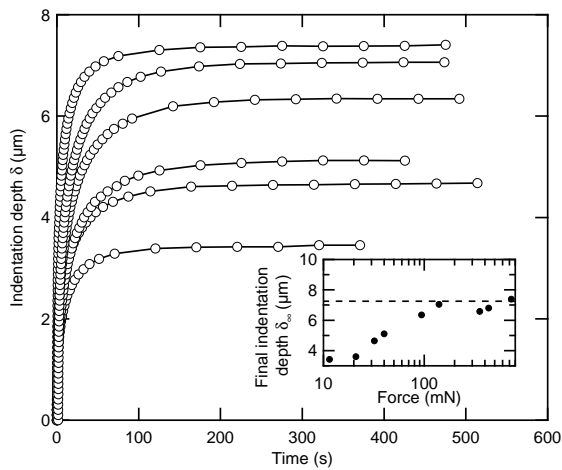}
		\caption{Measured indentation depth $\delta$ of a poly(PEGMA) film (swollen thickness: $e_0=9~\mu$m) as a function of time for increasing applied loads (radius of curvature of the probe: 5.2 mm). Applied loads from bottom to top : 11.5, 32, 40, 94, 140 and 730~mN. Inset : changes in the equilibrium indentation depth $\delta{_\infty}$ as a function of the applied load. Solid lines are guides for the eye.
		} 
		\label{fig:drainage}
	\end{center}
\end{figure}

\begin{figure}
	\begin{center}
		\includegraphics[width=0.9\columnwidth]{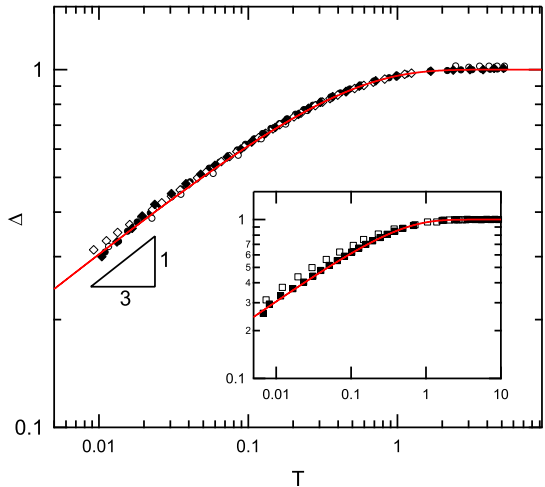}
		\caption{(Color on line) log-log plot of the normalized indentation depth $\Delta=\delta/\delta_\infty$ versus the non dimensional time $T=t/\tau$ for various contact conditions (film thickness: $e_0=9\:\mu$m). ($\circ$) $R=20.7$~mm, $F=$~7~mN; ($\lozenge$) $R=20.7$~mm, $F=$~88~mN; ($\bullet$) $R=5.2$~mm, $F=$~32~mN; ($\blacklozenge$) $R=5.2$~mm, $F=$~11.5~mN. For each data set, $\Delta$ and $\tau$ were obtained from a fit to eq.~(\ref{eq:sol}). Inset: same with $e_0=5\:\mu$m and ($\square$) $R=20.7$~mm, $F$=25~mN; ($\blacksquare$) $R=5.2$~mm, $F=$~20~mN. Red lines: variations of $\Delta$ with $T$ from eq.~(\ref{eq:sol}).}
		\label{fig:fit_PEGMA}
	\end{center}
\end{figure}

\begin{figure}
	\begin{center}
		\includegraphics[width=0.9\columnwidth]{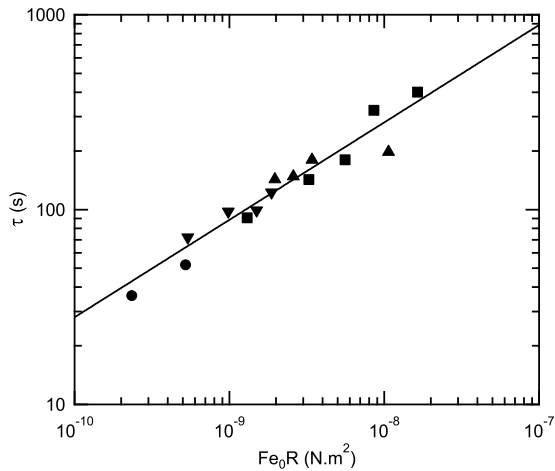}
		\caption{Characteristic poroelastic time $\tau$ versus $Fe_0R$. ($\bullet$)~$e_0=5\:\mu$m, $R=5.2$~mm ($\blacksquare$) $e_0=9\:\mu$m, $R=20.7$~mm; ($\blacktriangledown$)~$e_0=9\:\mu$m, $R=5.2$~mm; ($\blacktriangle$)~$e_0=5\:\mu$m, $R=20.7$~mm. The solid line corresponds to a power law fit with the exponent $\beta=0.52\pm0.06$.}
		\label{fig:tau_PEGMA}
	\end{center}
\end{figure}

We first consider the indentation response of poly(PEGMA) films for which neither glass transition nor phase separation transition is expected to occur during the course of poroelastic drainage. Fig.~\ref{fig:drainage} shows a typical example of the measured time dependence of indentation kinetics under increasing applied normal loads for a water swollen film of thickness $9\:\mu$m in contact with a glass probe of radius $5.2$~mm. Within a few tens of seconds, the indentation depth progressively reaches a limiting value $\delta_\infty$ corresponding to highly confined contact conditions ($\sqrt{R\delta_\infty} /e_0>>10$). As expected, the final indentation depth increases with the applied load as more and more water is expelled from the hydrogel film. However, inset in Fig.~\ref{fig:drainage} indicates that $\delta_\infty$ saturates at a maximum value $\delta_{max}$ of about $7\:\mu$m when the applied normal load is increased.
This maximum value of the indentation depth can be compared to the indentation depth $\delta_{dry}$ which would correspond to a complete drying of the films during the course of poroelastic indentation. A lower bound of $\delta_{dry}$ which neglects the indentation depth of the film in the drained state can simply be expressed in the form
\begin{equation}
	\delta_{dry}=e_0-e_{dry}=e_{dry}(S-1)
	\label{eq:indentation_dry}
\end{equation}
where $e_0$ and $e_{dry}$ are the initial and dry film thickness, respectively and $S$ is the film swelling ratio. For the film thickness considered in Fig.~\ref{fig:drainage}, this calculation yields $\delta_{dry}=5.7\:\mu$m, i.e. a value less than the measured value of $\delta_{max}$. Taking into account that $\delta_{dry}$ corresponds to a lower bound which does not account for the indentation of the fully dried film, this result indicates that the maximum value of the indentation depth of the poly(PEGMA) film corresponds to a fully drained state of the polymer network. The glass probe is even found to indent the rubbery dry polymer.\\
\begin{figure}
	\begin{center}
		\includegraphics[width=0.9\columnwidth]{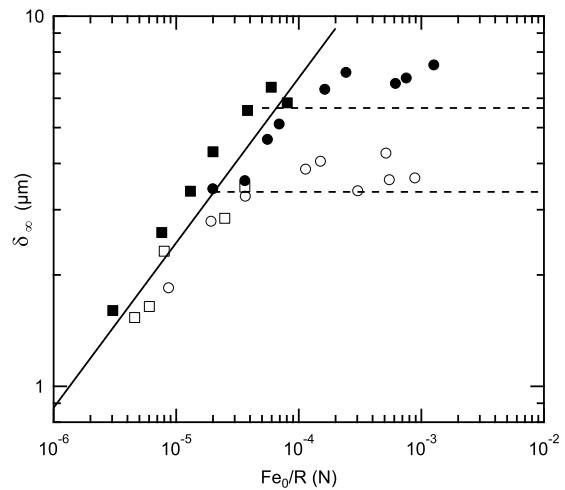}
		\caption{Equilibrium indentation depth $\delta_\infty$ versus reduced applied load ${Fe_0}/{R}$ for poly(PEGMA) films. Horizontal dashed lines corresponds to the calculated (eq.~(\ref{eq:indentation_dry})) maximum indentation depth $\delta_{dry}$ for $e_0=5\:\mu$m (lower) and $e_0=9\:\mu$m (upper). For probe radius $R=20.7 $~mm: ($\square$) $e_0=5\:\mu$m, ($\blacksquare$) $e_0=9\:\mu$m; for probe radius $R=5.2$~mm: ($\circ$) $e_0=5\:\mu$m, ($\bullet$)  $e_0=9\:\mu$m. The solid line corresponds to a power law fit (with the exponent $\alpha=0.44 \pm 0.09$) of the all data points verifying $\delta_\infty  <\delta_{dry}$.}
		\label{fig:delta_inf_PEGMA}
	\end{center}
\end{figure}
Separate lateral contact experiments were carried out in order to investigate potential changes in the shear properties of the drained poly(PEGMA) films. Preliminary measurements in air using dry films yielded a value of 60 kPa for the storage shear modulus  which is consistent with previously reported bulk values of similar poly(PEGMA) networks.~\cite{chan2012b,bryan2001} No significant change in the measured lateral contact stiffness of poly(PEGMA) films was found to occur during the course of poroelastic indentation of fully immersed contacts, even under the highest contact loads where $\delta_{\infty}=\delta_{\max}$ (results not shown). From this observation it can be concluded that the shear modulus of the poly(PEGMA) film remains constant whatever its water content. Such behaviour is consistent with the fact that no glass transition or phase separation of the poly(PEGMA) network is expected to occur when the swollen hydrogel network is dried out. It also complies with the hypothesis of a constant shear modulus $G$ of the network which is embedded in our contact model.\\
The validity of the indentation model for poly(PEGMA) films was further examined from a systematic investigation of the equilibrium indentation values $\delta_\infty$ as a function of film thickness $e_0$, probe radius $R$ and applied load $F$. As suggested by eq.~(\ref{eq:delta_inf}), experimental $\delta_\infty$ values have been reported in Fig.~\ref{fig:delta_inf_PEGMA} as a function of $Fe_0/R$. In this figure, horizontal dashed lines correspond to the maximum indentation depths $\delta_{dry}$ calculated from eq.~(\ref{eq:indentation_dry}) for the two film thicknesses under consideration. The measured maximum values of the indentation depth under the highest values of $Fe_0/R$ are slightly above these limits, thus indicating a fully drained state of the film. Consistently with the theoretical prediction (eq.~(\ref{eq:delta_inf})), all the data verifying $\delta_\infty < \delta_{max}$ follow a power law dependence on $Fe_0/R$ with an exponent ($\alpha=0.44 \pm 0.09$) which is reasonably close to the theoretical value ($\alpha=0.5$, cf eq.~(\ref{eq:delta_inf})) if one consider the experimental scatter in the data.\\
In order to obtain the characteristic poroelastic time $\tau$, indentation kinetics were also systematically fitted using eq.~(\ref{eq:sol}) with $\tau$ and $\delta_\infty$ as fitting parameters. Some examples of the resulting fits have been reported in a non dimensional form in the log-log plot in Fig.~\ref{fig:fit_PEGMA}. This figure as well as the inset shows that experimental data collected with two different thicknesses $e_0$, varied radii of curvature and applied loads all fall on a single master curve. It therefore emerges that our approximate contact model is able to predict accurately the indentation kinetics of poly(PEGMA) films over the considered range of film thickness and contact conditions. The characteristic times $\tau$ extracted from the fits are reported in a log-log plot as a function of ${Fe_0R}$ in Fig.~\ref{fig:tau_PEGMA}. As expected from eq.~(\ref{eq:tau}), they obey a power law dependence on ${Fe_0R}$ with an exponent $\beta=0.52\pm0.06$ which is very close to the theoretical prediction of 0.5.\\
In the case of rubbery poly(PEGMA) films, the above developed approximate poroelastic model is therefore found to provide a consistent description of the poroelastic indentation kinetics. If very precise measurements of the poroelastic time, diffusion coefficient or longitudinal modulus were required, one could argue that previously developed numerical models would probably provide more accurate data than the present approximate model, especially in the case of moderately confined contact geometries. However, our model circumvents the complexities of Finite Element approaches to provide in the form of simple scaling laws the dependence of equilibrium indentation depth and poroleastic time on contact parameters and film properties. Moreover, we show below that it is sensitive to the occurrence of glass transition or demixing transitions which is the objective of this investigation rather than a precise measurement of poroelastic time and subsequent material property values which was already addressed in previous studies. 
\section*{Poroelastic response of poly(NIPAM) and poly(DMA) films}
\begin{figure}
	\begin{center}
		\includegraphics[width=0.9\columnwidth]{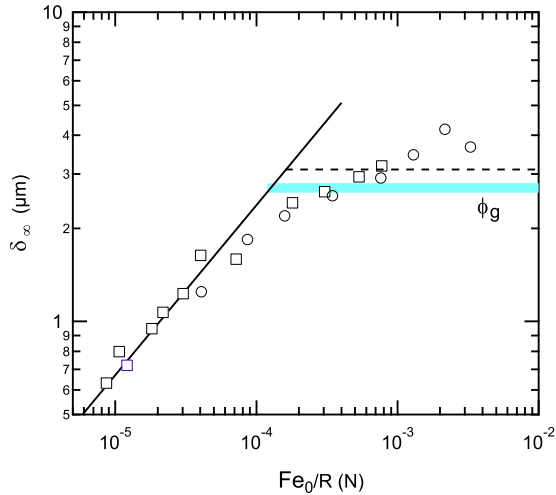}
		\caption{(Color on line) Equilibrium indentation depth $\delta_\infty$ versus reduced applied load ${Fe_0}/{R}$ for poly(DMA) films (thickness $e_0=4.5\:\mu$m). ($\circ$) $R=5.2 $~mm; ($\square$) $R=20.7 $~mm. Horizontal dashed lines corresponds to the calculated (eq.~(\ref{eq:indentation_dry})) maximum indentation depth $\delta_{dry}$. The horizontal blue line indicates the indentation depth corresponding to the water volume fraction $\phi_g=16$~vol\% at which the glass transition is expected to occur. The solid line corresponds to a power law fit (with the exponent $\alpha=0.55 \pm 0.03$) of the data points verifying $\delta_\infty <2\:\mu$m .}
		\label{fig:delta_inf_PDMA}
	\end{center}
\end{figure}

\begin{figure}[h]
	\begin{center}
		\includegraphics[width=1\columnwidth]{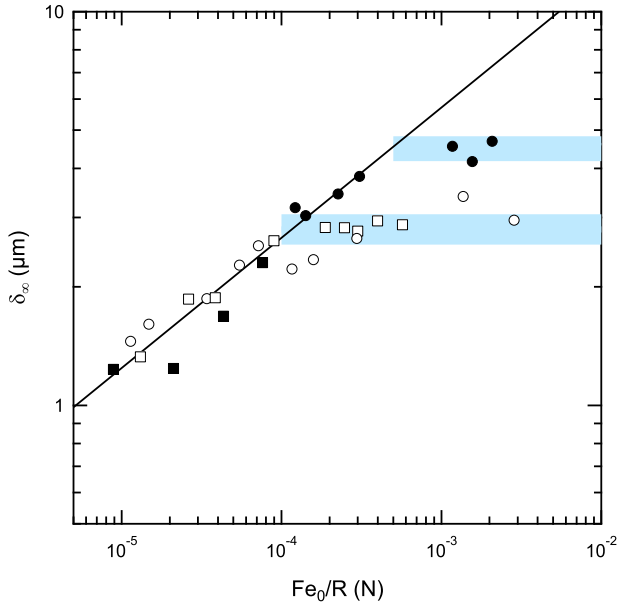}
		\caption{Equilibrium indentation depth $\delta_\infty$ versus reduced applied load $Fe_0/R$ for poly(NIPAM) films. Probe radius $R=5.2$~mm: ($\circ$)~$e_0=5.9\:\mu$m; ($\bullet$)~$e_0=11.5\:\mu$m. $R=20.7$~mm: ($\square$)~$e_0=5.9\:\mu$m; ($\blacksquare$)~$e_0=11.5\:\mu$m. The solid line corresponds to a power law fit (exponent $\alpha=0.33 \pm 0.02$) of the data points verifying $\delta_\infty  <\delta_{max}$ where $\delta_{max}$ is the measured maximum value of the indentation depth. Blue areas delimit indentation depths corresponding to water volume fractions of $\phi=0.40 \pm 0.05$ (lower, $e_0=5.9\mu$m) and $\phi=0.50 \pm 0.05$ (upper, $e_0=11.5\mu$m).}
		\label{fig:delta_inf_PNIPAM}
	\end{center}
\end{figure}

\begin{figure}[h]
	\begin{center}
		\includegraphics[width=1\columnwidth]{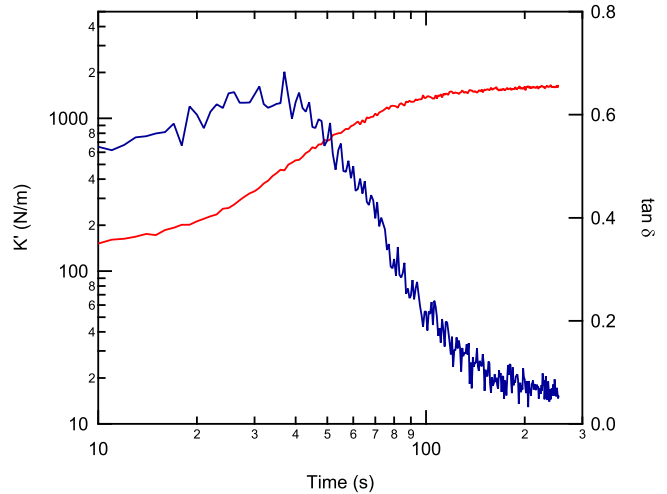}
		\caption{(Color online) In-phase lateral contact stiffness $K'$ (red) and damping factor $\tan\:\delta$ (blue) versus time for a poly(NIPAM) film of thickness $e_0=4.8\:\mu$m (applied laod: 2~N).}
		\label{fig:lateral_PNIPAM}
	\end{center}
\end{figure}

\begin{figure}[h]
	\begin{center}
		\includegraphics[width=0.6\columnwidth]{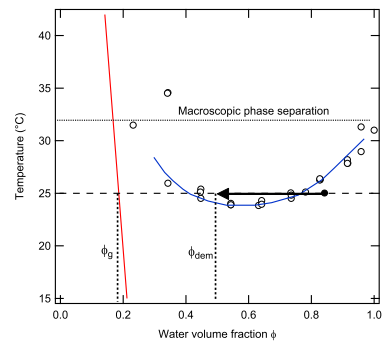}
		\caption{(Color on line) Schematic of the temperature composition phase diagram of the poly(NIPAM)/water system showing both the glass transition (red line, as predicted from eq.~(\ref{eq:fox})) and the phase separation (blue line, taken from data reported by Van Durme~\textit{et~al}~\cite{vandurme2004} boundaries. The black arrow corresponds to the hypothetical path followed by confined films during poroelastic indentation.}
		\label{fig:phase_diagram}
	\end{center}
\end{figure}

Fig.~\ref{fig:delta_inf_PDMA} shows the measured value of $\delta_{\infty}$ as a function of $Fe_0/R$ for a poly(DMA) film $4.5\:\mu$m in thickness in contact with two different spherical probes ($R=$~5.2 and 20.7~mm). As for poly(PEGMA) films, the maximum values $\delta_{max}$ of the measured indentation depths are slightly above the calculated $\delta_{dry}$ value (horizontal dashed line in Fig.~\ref{fig:delta_inf_PDMA}), thus indicating a nearly fully dried state of the film under the highest loads. However, a departure from the power law dependence can be noted for $\delta_{\infty}$ values about half of the measured maximum indentation depth.\\
\begin{figure}[h]
	\begin{center}
		\includegraphics[width=0.8\columnwidth]{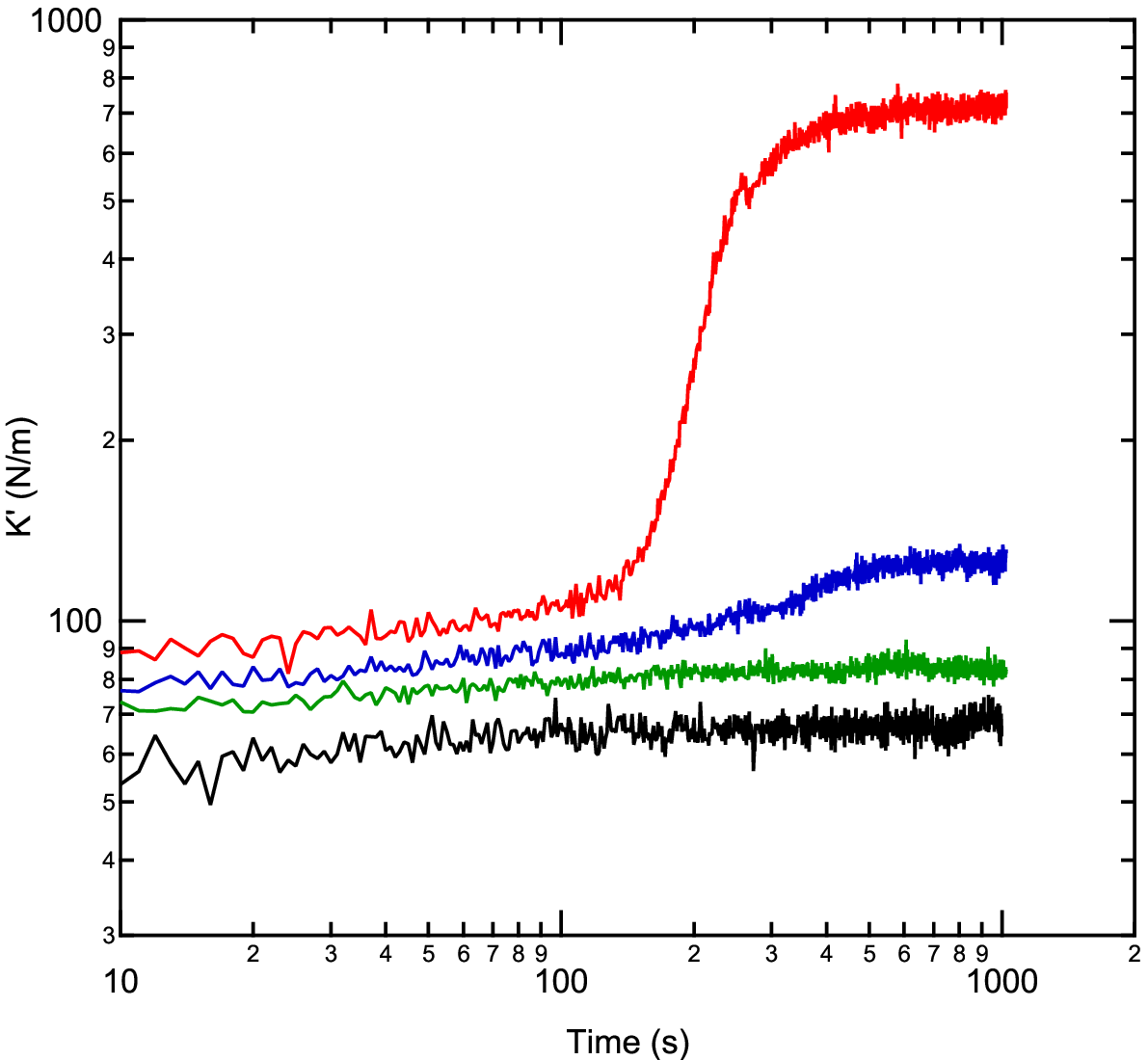}
		\includegraphics[width=0.8\columnwidth]{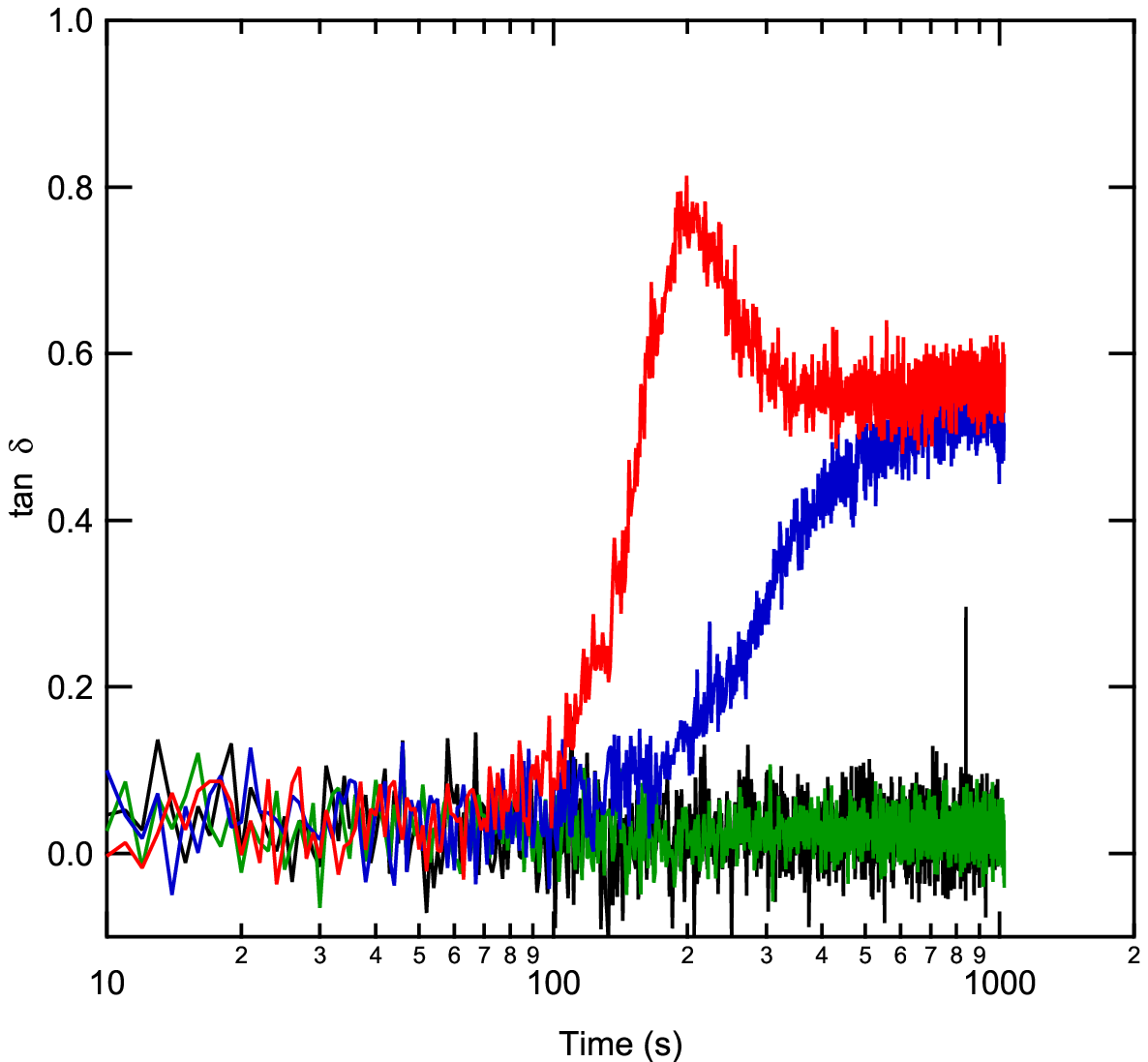}
		\caption{(Color online) In-phase lateral contact stiffness $K'$ (top) and damping factor $\tan\:\delta$ (bottom) versus time for a poly(DMA) film of thickness $e_0=4.7\:\mu$m. Applied load $F$: (black) 1 N; (green) 2 N; (blue) 3 N; (red) 4 N.}
		\label{fig:lateral_PDMA}
	\end{center}
\end{figure}
Interestingly, lateral contact stiffness measurements carried out in the domain where $\delta_{\infty}$ data deviate from the expected power law dependence reveal the occurrence of strong changes in the mechanical properties of the films. Fig.~\ref{fig:lateral_PDMA} illustrates the time dependence of the in-phase lateral contact stiffness $K'$ and in the damping factor $\tan \delta$ for increasing applied loads corresponding to $Fe_0/R$ values in the range 0.9~$10^{-4}$ - 7.6~$10^{-4}$. Within this domain, increasing the contact load results in a time-dependent increase in $K'$ up to nearly one order in magnitude. At the same time, $\tan \delta$ is also increasing, even showing a peak at the highest load (4~N). The time scale of these changes in the shear properties is reduced when the load is increased. It is also found to correspond to the characteristic poroelastic times measured during indentation experiments under similar contact conditions.\\
These observations can be attributed to the fact that the water plasticized poly(DMA) film is progressively reaching its glass transition domain as the water is expelled out. According to Fox equation~\cite{fox1956}, the glass transition temperature of the water/polymer mixture $Tg_{wp}$ can be expressed as
\begin{equation}
	\frac{1}{Tg_{wp}}=\frac{1-w}{Tg_{p}}+\frac{w}{Tg_{w}}
	\label{eq:fox}
\end{equation}
where $Tg_{w}$ and $Tg_{p}$ are respectively the glass transition of water (-137$^\circ C$) and poly(DMA) (107$^\circ C$ as measured using Differential Scanning Calorimetry) and $w$ is the weight fraction of water. Following this approach, the glass transition of the water/poly(DMA) system is expected to occur at room temperature for a water volume fraction of $\phi_g=16$~vol\% which, as indicated by the horizontal blue line in Fig.~\ref{fig:delta_inf_PDMA}, corresponds to an indentation depth of about 2.7~$\mu$m. This indentation depth lies in the range where both a departure from the power law dependence of $\delta_{\infty}$ on $Fe_0/R$ (Fig.~\ref{fig:delta_inf_PDMA}) and an increase in the shear stiffness of the film (Fig.~\ref{fig:lateral_PDMA}) are observed. These observations support the hypothesis of a poroelastic indentation depth limited by the glass transition of the water plasticized poly(DMA) gel.\\
When the poroelastic response of poly(NIPAM) films is considered, one of the most striking difference with both poly(PEGMA) and poly(DMA) films is that the measured values of the maximum indentation depth are well below the computed indentation depth which would correspond to a fully drained state i.e. $\phi \sim 0$. As indicated by the blue areas in Fig.~\ref{fig:delta_inf_PNIPAM}, the measured maximum indentation depths for $e_0=$~5.9 and 11.5~$\mu$m correspond to water volume fractions of $\phi=0.50 \pm 0.05$ and $\phi=0.40 \pm 0.05$, respectively. The films are thus far to be fully drained even under the highest loads. In addition, lateral contact experiments under a normal load of 2~N for a film $4.8\:\mu$m in thickness (i.e. $Fe_0/R=3.9\:10^{-4}$~N) show a clear transition in the shear properties of the film during the course of poroelastic indentation (Fig.~\ref{fig:lateral_PNIPAM}) with a damping peak and a nearly one order of magnitude increase in the contact stiffness $K'$. The maximum in $\tan \delta$ occurs at $t=40$~s which lies within the time scale of poroelastic indentation experiments carried out under similar contact conditions.\\
As for poly(DMA) films, one could tentatively attribute these changes in shear properties to the glass transition of the water plasticized poly(NIPAM) polymer. At an ambient temperature of 25$^\circ C$, the water volume fraction $\phi_g$ can be equivalently estimated either using Fox equation (eqn~(\ref{eq:fox})) with $Tg_p=130^\circ C$ for poly(NIPAM)): $\phi_g=18$~vol\%, or from the literature, where experimental values were obtained from Differential Scanning Calorimetry (DSC) measurements: $\phi_g=18$~vol\% measured at 22$^\circ C$ by Afroze~\textit{et al}?~\cite{afroze2000} for linear poly(NIPAM). Such a $\phi_g$ value is much less than the measured minimum water volume fraction within the indented films ($\phi \sim$~0.4-0.5) as indicated in  Fig.~\ref{fig:delta_inf_PNIPAM}, thus allowing us to discard the hypothesis of an indentation depth limited by the glass transition of the network.\\
Alternately, the observed changes in shear properties could be attributed to the widely reported phase separation or demixing transition of poly(NIPAM) systems.~\cite{shibayama1993,cai2011} Poly(NIPAM)/water mixtures are known to exhibit solubility that depends both on the water content and on the temperature. The phase separation behavior is often associated to a coil-to-globule transition of the polymer chains under the action of hydrogen bonding and hydrophobic interactions.~\cite{schild1992} It is characterized by a lower critical solution temperature (LCST) which reported value\cite{halperin2015} varies between $24^\circ$~C and $28^\circ$~C depending on investigators and a critical water volume fraction of $60\pm$5~vol\%. These critical temperature and composition are reported to be independent of the polymer molar mass and are unchanged for a polymer network.~\cite{afroze2000,vandurme2004}\\
For a network of cross-linked poly(NIPAM), the swelling at equilibrium results from a balance between the mixing free energy and the elastic contribution of the polymer network. The former term depends on the solubility parameter. Its dependence with composition and temperature is believed to result in the existence of a three phase equilibrium between pure solvent, a dilute, and a concentrated network phase, the latter two phases remaining mixed at small scale within a given range of temperature and composition.~\cite{afroze2000} It further leads to a macroscopic volume transition reported at temperatures above 32-35$^\circ$C for which bulk gels of poly(NIPAM) abruptly collapse.~\cite{shibayama1993}  If such a macroscopic transition was to happen in our case, a sharp increase in the indentation depth would be observed. We therefore emphasize that no macroscopic phase separation occurs during the measurement period.\\
On the other hand, changes in the elastic term in the free energy of swollen polymer networks also modify the poly(NIPAM) phase behavior. Indeed, the phase separation behavior of poly(NIPAM) is also known to be influenced by mechanical stresses. Various studies report on volume-phase transition of \textit{N}-isopropylacrylamide induced at 20$^\circ C$ by hydrostatic pressure.~\cite{kato1997,Nasimova2004} However, such a pressure-induced transition is not likely to occur in our experiments as it involves hydrostatic pressures (about 200 MPa at 25$^\circ C$) one order of magnitude higher than the contact pressures under consideration (no more than 10~MPa). More interestingly, stretching is known to induce an increase in the LCST of poly(NIPAM)/water networks.~\cite{suzuki1997,suziki1994} Conversely, a decrease in the LCST could be anticipated under the action of a compressive stress. Finally, as reported by Ilavsky,~\cite{ilavsky1993} the phase transition of acrylamide gels can result in a one order of magnitude increase in the shear modulus which is consistent with the present $K'$ measurements for poly(NIPAM) gels.\\
As schematically depicted in Fig.~\ref{fig:phase_diagram}, a possible explanation for the observed poroelastic indentation behavior of poly(NIPAM) networks at ambiant temperature (25$^\circ$C) would be that, in a temperature-composition phase diagram, the swollen network approaches the critical point or even crosses the demixing boundary (shown as a blue line based on the data by Van Durme~\textit{et al})~\cite{vandurme2004} as water is expelled out and so, before the glass transition boundary (red line in Fig.~\ref{fig:phase_diagram}). Such an hypothesis could account for the observation that the measured increase in shear properties and the associated limitation in $\delta_{\infty}$ occur at a water volume fraction $\phi\sim$0.4 to 0.5 well above $\phi_g$, in the range where demixing has been observed in the literature.\\
Our experimental results suggest that under isothermal conditions, expelling water out of the poly(NIPAM) network causes an arrested demixing transition at a molecular scale, without a macroscopic phase separation during the measurement period. Although poorly understood, the occurrence of such arrested demixing is reported in the recent review by Halperin~\textit{et al}~\cite{halperin2015} where it is tentatively attributed to the formation of long-lived particles denoted as mesoglobules.\\
As a summary, it emerges that both the poly(DMA) and poly(NIPAM) confined films can experience strong changes in their shear properties during the course of poroelastic indentation as result of either glass transition or phase separation phenomena. These transitions result in deviations from the theoretical predictions of the equilibrium indentation depth as a function of film thickness and contact conditions. The magnitude of the changes in the shear properties clearly precludes any attempt to apply the above developed poroelastic indentation model which relies on the assumption that the shear modulus is independent on water concentration.In addition, one could anticipate that the diffusion coefficient is also strongly dependent on phase transition. A complete description of the indentation behaviour of confined gel films in such situations would require a coupled poroelastic model able to handle the water concentration dependence of both the shear modulus and the diffusion coefficient. Rather than a scaling approach, such a description would imply a complete reformulation of the problem within the framework of poroelasticiy which is beyond the scope of this study. Moreover, the derivation of a theoretical description of the water concentration dependence of the shear modulus can be anticipated to be a major issue depending on the nature of the transition. While in the case of glass transition it could probably be handled through well-established Fox-Flory equation and time-temperature superposition principle, a theoretical description of this phenomena remains an open issue in the case of demixing transitions such as these encountered for poly(NIPAM) films.
\section*{Conclusion}
In this work, we have investigated the poroelastic indentation behavior of hydrogel thin films (less than 15~$\mu$m) mechanically confined between rigid glass substrates. Different types of polymer networks have been considered which differ in their physical properties in the dry state. While the dry poly(PEGMA) network is rubbery at room temperature, both poly(DMA) and poly(NIPAM) networks are glassy polymers in the dry state. In addition, poly(NIPAM) is also known to experience a demixing transition (LCST) under the action of temperature, composition or pressure changes. In the case of the poly(PEGMA) films, we have shown that the enhanced contact pressures which result from the geometrical confinement of the films can induce a complete drainage of the network down to a vanishing water content. Even under such large variations in water content, no significant changes in the shear properties of the films are evidenced from lateral contact stiffness measurements. The indentation kinetics can be satisfactorily described over the whole range of contact conditions and film thickness using an analytical poroelastic contact model developed within the limits of confined contact geometries. This approximate model allows to predict the equilibrium indentation depth and the poroelastic time in the form of simple scaling laws of contact parameters and film thickness, with the network properties (permeability, shear modulus, Poisson's ratio) and the fluid viscosity as parameters.\\ 
Conversely, deviations from this model are evidenced for poly(DMA) and poly(NIPAM) system for which the maximum achievable indentation depth is strongly reduced: drainage is found to be blocked below some threshold in the water volume fraction which, on the other hand, corresponds to a one order of magnitude increase in the shear properties of both poly(DMA) and poly(NIPAM) films. In the case of poly(DMA) system, this threshold is found to be associated with the glass transition of the water plasticized polymer network. For poly(NIPAM), it turns out that the observed transition is rather related to an arrested demixing transition inducing a phase separation at molecular scale.\\
These results emphasize the necessity of carefully considering the details of physical chemistry of hydrogel networks under such confined contact conditions or whenever the water content is reduced enough for drastic changes in the polymer network to happen, a point often overlooked in the literature. Further implications may be expected regarding the frictional properties of thin hydrogel films which could exhibit similar transitions depending on the residence time within sliding contacts.

\section*{Acknowledgments}
One of us (J. Delavoipi\`ere) is indebted to Ekkachai Martwong for his kind support during the synthesis of the hydrogels systems. The authors also wish to thank C.-Y. Hui for stimulating discussions.
%
\bibliographystyle{unsrt} 

%
\end{document}